\documentstyle[prl,aps,amsmath,amssymb]{revtex}
\newcommand{\be}{\begin{eqnarray}}
\newcommand{\ee}{\end{eqnarray}}

\newcommand{\bi}{\begin{itemize}}
\newcommand{\ei}{\end{itemize}}
\begin{document}
\twocolumn[\hsize\textwidth\columnwidth\hsize
           \csname @twocolumnfalse\endcsname
\title{Bifurcation in kinetic equation for interacting Fermi systems}
\author{Klaus Morawetz}
\address{Max-Planck-Institute for the Physics of Complex Systems, 
Noethnitzer Str. 38, 01187 Dresden, Germany}
\maketitle
\begin{abstract}
The recently derived nonlocal quantum kinetic equation for dense
interacting Fermi
systems  
combines time derivatives with finite time stepping known from
the logistic mapping. This  continuous delay differential equation 
equation is a consequence of the microscopic delay time representing
the dynamics of the deterministic chaotic system. 
The responsible delay time is explicitly
calculated 
and discussed for short range correlations. As a novel feature 
oscillations in
the time evolution of the distribution function itself
appear and bifurcations
up to chaotic behavior occur. The temperature and density conditions
are presented where such oscillations and bifurcations arise
indicating an onset of phase transition.
\end{abstract}
\pacs{82.40.Bj,05.20.Dd,05.70.Ln,82.20.Mj}
\vskip2pc]

{\bf The relation between microscopic chaos and kinetic theory within the time evolution of averaged distribution functions is a matter of ongoing debate. If the coarse graining of phase space due to averaging is large enough, any microscopic signal of chaoticity like bifurcation is lost. The time evolution of the distribution function evolves continously and smoothly in time. If the coarse graining is reduced by considering more microscopic fluctuations the time evolution of the distribution function itself can reveal signs of the underlying chaotic motion. To understand this relation different attempts have been made in the literature starting from standard mapping and deriving an appropriate kinetic description in terms of finite time step equations. Here the opposite route is proposed: Starting from a quantum statistical approach we employ the nonlocal kinetic theory for dense Fermi systems to show that indeed the resulting kinetic equation can exhibit a nonmonotonic time evolution in a specific temperature and density range. This is a consequence of the considered many-body correlations. Within a first approximation of relaxation times we show that a delay differential equation appears which interpolates between finite time stepping like logistic mapping and continuous time evolution by differential operators. It is argued that the occurring bifurcations in the time evolution are a signal of onsetting phase transition.}

\section{Introduction}
The kinetic theory of interacting quantum or classical 
gases and the description of
deterministic chaotic systems are mostly developed separately with very
few overlapping. Almost exclusively for the case of Lorentz gas the interlink
has been analyzed \cite{KRN00,DP97}, and citation therein. The kinetic
theory  understands the increase of entropy as a result of many random
collisions. The theory of deterministic chaos on the other hand
represents the irreversibility by the characteristic measure of
Lyapunov exponent. While the kinetic theory can be easily extended to
quantum systems the quantum chaos is still a matter of debate about
even the correct definition. 

If both approaches describe some facet of irreversibility it should be
possible to give relations between them. For instance, one can connect
the transport coefficients with the Lyapunov exponent 
\cite{evans,dorfman,dorfman1,C95,DB97,M99}. In \cite{evans,C95} the fact, that
the spreading of a small phase space volume is given by the sum of
Lyapunov exponents,
is used to give a relation between Lyapunov exponents and
viscosity. In \cite{dorfman,dorfman1} the relation between transport
coefficients and Lyapunov exponents was presented in terms of
Helfand's moments. The interlink was possible to establish by
reinterpretation of the Helfand's moments as stochastic quantities
such that the mean variance of the time derivatives represents just
the transport coefficients. In \cite{DB97} the authors derive a
density expansion of the largest Lyapunov exponent for hard sphere gases
from a generalized Lorentz-Boltzmann equation. The intimate
relation 
between transport coefficients and dynamical
quantities like the Lyapunov exponent is even apparent in quantum Fermi
gases in that an additional chaotic process behaves like an additional relaxation time \cite{M99}. 

Recently another more direct relation between kinetic theory and
deterministic chaotic systems has been established \cite{Ba00}. It
appears for
the standard mapping that the phase-space averaged kinetic description can be
formulated and solved in terms of
finite time step equations instead of differential ones. We want to
address here the opposite route starting from a continuous kinetic equation
for strongly correlated quantum Fermi systems and will ask whether there
is a reminiscence of the underlying chaotic motion in the time
evolution of the distribution function itself. For a system of
hard-sphere gases the appropriate kinetic description is the Enskog
equation consisting of nonlocal off-sets in the collision
integral. Here we will consider the opposite extreme of particles
interacting in such a way that they form short-living molecules as
correlated states. 
We will show here that such collision
duration leads to delay
differential equations.  
In particular,
it will be found that the time
evolution of the distribution function itself undergoes
oscillations up to chaotic
behavior similar to Hopf bifurcations observed in mappings including
memory \cite{HF93}. This is due to a competition between 
attractive potential and
repulsive Pauli-blocking.  It can be
considered as a dynamical signal of the onset of phase transition due to the 
formal similarity between mean-field phase transitions and
bifurcations of one-dimensional discrete maps \cite{FKR87}.

The understanding of the route to chaos has
been paved with the logistic mapping \cite{F78}
\be
x_{k+1}=a x_k (1-x_k)
\label{1}
\ee
which starts to show successive bifurcations for increasing parameters
$3\le a$, the chaos occurring at $a=3.5699..$. 
An interesting extension of this
logistic model has been studied by Berezowski \cite{B01} in that a
finite inertia has been added
\be
\sigma {d f(t) \over d t} +f(t)=a f(t-\tau_d) [1-f(t-\tau_d)]
\label{2}
\ee
with a delay time $\tau_d$. For $\sigma=0$ this model reduces to the
logistic mapping (\ref{1}). In general such retarded differential
equations (\ref{2}) are very complex and are
used to model periodic and aperiodic dynamics of physiological 
systems \cite{MG77}. It is possible to simulate nonlinear dynamical
systems by linear delay-differential equations in loading information
in the initial conditions \cite{J00}. 

As done in \cite{B01} we perform a a linear stability analysis of
(\ref{2}) according to $f=f_0+\delta f$ with the fixed point solution of $-f_0+a f_0 -a f_0^2=0$. Without loss of generality, we look only at the 
instable one
%\be
$f_0={a-1\over a}$.
%\ee
The Fourier-transformed disturbance  
%\be
$\delta f(t) =\int {d \omega \over 2 \pi} {\rm e}^{-i \omega t} f(\omega)$
%\ee
determines then the complex frequency $\omega=\Omega+i \Gamma$ which 
gives for $\Gamma>0$ the instable and for $\Gamma<0$ the stable solutions.
One gets from (\ref{2})
\be
-i\omega \sigma +1 -a {\rm e}^{i \omega \tau_d} (1-f_0)=0.
\label{eq}
\ee
Separating real and imaginary part of (\ref{eq}) one has
\be
1+\Gamma \sigma &=& {\rm e}^{-\Gamma \tau_d} (2-a) \cos \Omega
\tau_d\nonumber\\
-\Omega \sigma &=& {\rm e}^{-\Gamma \tau_d} (2-a) \sin \Omega \tau_d.
\label{2a}
\ee
Dividing both sides of (\ref{2a}) gives the equation for the frequency
$\Omega$
\be
\Omega \sigma +(1+\Gamma \sigma) \tan \Omega \tau_d=0
\label{om1}
\ee
and adding both squared sides of (\ref{2a}) gives
\be
\sqrt{(1+\Gamma \sigma)^2 +\Omega^2 \sigma^2}={\rm e}^{-\Gamma \tau_d
  } |2-a|.
\label{3}
\ee
Now we look for instable modes $\Gamma>0$ which leads to  
\be
\sqrt{(1+\Gamma \sigma)^2 +\Omega^2 \sigma^2}\stackrel{<}{>}
|2-a|\qquad {\rm for}\quad \tau_d \stackrel{>}{<
}0.
\label{3}
\ee
In the final form we assume $\Gamma \sigma\ll1$ and we see
that the presence of the term $\sigma$ shifts
the onset of bifurcations and of chaos towards higher (lower) values of
$a$ if compared to the logistic map
\be
|a-2|\stackrel{>}{<} \sqrt{1+(\sigma \Omega)^2}
\label{crit}
\ee
for positive (negative) delay times $\tau_d$ respectively.
Furthermore an oscillation appears with a frequency $\Omega$ 
determined by (\ref{om1})
\be
\tau_d \Omega +\arctan (\sigma \Omega)=\pi.
\label{om}
\ee
There exists a limiting value for the delay time $\tau_d$ below
which all oscillations dissappear. These oscillations and the finite
delay time $\tau_d$ are the reasons why  Hopf
bifurcations can occur in the model (\ref{2}) analogously to the model
considered in
\cite{HF93} where an additional memory has been added to (\ref{1}).

\section{Nonlocal kinetic theory}
It is now noteworthy to see that the model (\ref{2}) can in fact be
derived from microscopic quantum statistics and as such bears direct
physical
relevance. We will show that indeed for an interacting Fermi system at
low temperatures the one-particle distribution function obeys a
kinetic equation which can be written in the form of (\ref{2}). In this
way we 
will express the parameters $a$ and $\sigma$ by
characteristic  physical quantities of the system, the temperature $T$,
density $n$, scattering length $a_0$ and range of the potential $r_0$.

First let us recall the quantum nonlocal kinetic equation
for the quasiparticle distribution $f$ \cite{SLM96,LSM97} 
with quasiparticle
energy $\varepsilon$
\begin{eqnarray}
&&{\partial f_1\over\partial t}+{\partial\varepsilon_1\over\partial k}
{\partial f_1\over\partial r}-{\partial\varepsilon_1\over\partial r}
{\partial f_1\over\partial k}
%\nonumber\\
%&&
=\int{dpdq\over(2\pi)^5\hbar^7}P_-
\nonumber\\
&&\times\Bigl[\bigl(1\!-\!f_1\bigr)\bigl(1\!-\!f_2^-\bigr)f_3^-f_4^--
f_1f_2^-\bigl(1\!-\!f_3^-\bigr)\bigl(1\!-\!f_4^-\bigr)\Bigr].
\label{kin}
\end{eqnarray}
The superscripts $-$ denote the signs of non-local corrections:
$f_1\equiv f(k,r,t)$, $f_2^-\equiv f(p,r\!-\!\Delta_2,t)$,
$f_3^-\equiv f(k\!-\!q\!-\!\Delta_K,r\!-\!\Delta_3,t\!-\!
\Delta_t)$, and $f_4^-\equiv f(p\!+\!q\!-\!\Delta_K,r\!-\!
\Delta_4,t\!-\!\Delta_t)$.  The first(second) part of the collision integral is called in(out)-scattering since it (de)populates the distribution $f_1(k,r,t)$. The time-space symmetries and particle-hole symmetries of (\ref{kin}) are discussed in \cite{SLM98}. The form (\ref{kin}) used in the present paper, shows explicitly the particle-hole symmetry since the out-scattering is the particle-hole mirror of the in-scattering. Let us here remark only that it can be given an equivalent form with explicit time-space symmetry \cite{SLM98}.
The scattering measure $P_-=|{\cal T}^r_-|^2 \delta (\varepsilon_k^-+\varepsilon_p^--\varepsilon_{k+q}^--\varepsilon_{p-q}^--\Delta_E)$ is given by the
modulus of the scattering T-matrix which is a complex quantity ${\cal T}^r=|{\cal T}^r| \exp i\phi$. The latter one can be found from scattering theory or experimental phase shift analysis.
All corrections,
the $\Delta$'s, describing the non-local and non-instant collision \cite{SLM96,LSM97} are given by derivatives of the scattering phase shift
\mbox{$\phi={\rm Im\ ln}{\cal T}^r(\omega,k,p,q,t,r)$}
\begin{equation}
\begin{array}{lclrcl}\Delta_t&=&{\displaystyle
\left.{\partial\phi\over\partial\omega}
\right|_{\varepsilon_1+\varepsilon_2}}&\ \ \Delta_2&=&
{\displaystyle\left({\partial\phi\over\partial p}-
{\partial\phi\over\partial q}-{\partial\phi\over\partial k}
\right)_{\varepsilon_1+\varepsilon_2}}\\ &&&&&\\ \Delta_E&=&
{\displaystyle\left.-{1\over 2}{\partial\phi\over\partial t}
\right|_{\varepsilon_1+\varepsilon_2}}&\Delta_3&=&
{\displaystyle\left.-{\partial\phi\over\partial k}
\right|_{\varepsilon_1+\varepsilon_2}}\\ &&&&&\\ \Delta_K&=&
{\displaystyle\left.{1\over 2}{\partial\phi\over\partial r}
\right|_{\varepsilon_1+\varepsilon_2}}&\Delta_4&=&
{\displaystyle-\left({\partial\phi\over\partial k}+
{\partial\phi\over\partial q}\right)_{\varepsilon_1+\varepsilon_2}}.
\end{array}
\label{SHIFTSALL}
\end{equation}

The nonlocal kinetic equation (\ref{kin}) covers all quantum virial corrections on the
binary level and conserves density, momentum and energy including the
corresponding two-particle correlated parts \cite{LSM97}. The kinetic equation (\ref{kin}) allows for a simple classical interpretation of the complicated correlations covered by collisions. The collision integral can be understood as a collisional scenario that two particles are approaching until they reach their correlation distance $\Delta_2$ then they form a correlated pair with the collision duration of $\Delta_t$. During this correlated travel they can rotate so that they break up into single particles after the time $\Delta_t$ at the endpoints $\Delta_{3,4}$. 

The
classical kinetic theory
of dense gases of Enskog-like equations as well as the Landau theory
of interacting Fermi systems are limiting cases of this nonlocal
kinetic equation. It requires
no more technical problems than solving the Boltzmann equation \cite{MLSCN98,MLNCCT01}.
From the formal point of view this kinetic equation has been derived
taking into account all terms up to linear order in the quasiparticle
damping. 

We will now derive the equation (\ref{2}) from (\ref{kin}) and will
calculate the required parameters for a model ${\cal T}$-matrix of separable
potential.
First we analyze the collision integral and assume quadratic
dispersion relation for the quasiparticle energies. The 
$\delta$-function in ${\cal P}$ representing the energy conservation in the collision integral reads
\be
&&\delta({k^2\over 2 m_a}+{p^2\over 2 m_b}-{(k-q)^2\over 2 m_a}-{(p+q)^2\over 2 m_b})\nonumber\\
&&=\delta\left (|q|({k\over m_a}-{p\over m_b})\cdot {q\over
    |q|}-{q^2\over 2}
  ({1\over m_a} + {1\over m_b})\right )\nonumber\\
&&={\delta(|q|) \over \left ({k\over m_a}-{p\over m_b}\right)\cdot
    {q\over |q|}} + \delta(q\ne0)
\label{d}
\ee
with effective masses of particle $a,b$.
The second part is the usual considered $\delta$-function excluding zero transferred
momentum $q=0$. This collision part will be treated in
relaxation time approximation $\propto (F_0-f)/\tau$ where the
equilibrium distribution $F_0$ is specified later. 

More interesting is 
to notice that the
$q=0$ channel leads to an additional part absent in usual local kinetic
equations like the Boltzmann kinetic equation. 
To convince the reader of this novel observation let us rewrite the
Pauli-blocking factors of  (\ref{kin}) for the $q=0$ channel where
$f_3\rightarrow f_1$ and $f_4\rightarrow f_2$ according
to (\ref{d})
\be
&&\Bigl[\bigl(1\!-\!f_1\bigr)\bigl(1\!-\!f_2^-\bigr)f_3^-f_4^--
f_1f_2^-\bigl(1\!-\!f_3^-\bigr)\bigl(1\!-\!f_4^-\bigr)\Bigr]_{q=0}
\nonumber\\
&&=f_2^-(1-f_2^-) \Bigl[f_1^--f_1 \Bigr].
\ee
We see that for the case of local kinetic equations without delays
this specific channel disappears since $f_1^-=f_1$. Therefore we can
conclude that in
contrast to the usual local kinetic equation like the Boltzmann
equation, the nonlocal
equation possesses a finite zero angle channel in the collision
integral which is of course of mean-field type since no energy or
momenta is exchanged. We are now going to exploit this observation in
a specific way.

Since the signs of the shifts in the out-scattering parts are
arbitrary \cite{SLM98} we don't include them at all. Concentrating on the 
in-scattering
part we rewrite 
\be
&&f_3^- f_4^- (1-f_2^-)(1-f_1)|_{q=0}
\nonumber\\&&=
f_3^- (f_1^-\!-\!f_1)f_4^- (1\!-\!f_2^-)|_{q\!=\!0}\!+\!f_3^-
(1\!-\!f_1^-)
f_4^-
(1\!-\!f_2^-)|_{q\!=\!0}
\nonumber\\&&=
f_1^- (f_1^--f_1)f_2^- (1-f_2^-)+f_1^- (1-f_1^-)f_2^-
(1-f_2^-).
\nonumber\\&&
\label{rew}
\ee
The first part can be expanded 
\be
f_1^--f_1&=&-\Delta_t {\partial f \over \partial t}-\Delta_3 {\partial f
  \over \partial r}-\Delta_K {\partial f \over \partial k}\nonumber\\
&=&-\tilde\Delta_3 {\partial f
  \over \partial r}-\tilde\Delta_K {\partial f \over \partial k}
\label{form}
\ee
where in the last step we have replaced the time derivative of $f$ by
the free drift motion ${\partial f_1\over\partial t}=-{\partial\varepsilon_1\over\partial k}
{\partial f_1\over\partial r}+{\partial\varepsilon_1\over\partial r}
{\partial f_1\over\partial k}$ leading to derivative of on-shell
shifts
(\ref{SHIFTSALL}).
Therefore we can absorb the first part of (\ref{rew}) into the left
(drift) side of the nonlocal kinetic equation (\ref{kin}). These gradients are absent
for the 
now 
considered homogeneous case. Using thermal averaging of the occurring
microscopic quantities $\tau$, $\tilde\tau$ and $\Delta_t$ denoted by $<...>$ 
the second part of (\ref{rew}) leads to an additional term in the 
kinetic equation
\be
&&{d f(k,t)\over d t}=-{f(k,t)-F_0\over \tau}
+{f(k,t-\Delta_t)[1-f(k,t-\Delta_t)]\over \tilde \tau}
\nonumber\\&&
\label{kine}
\ee
with
\be
{1\over \tau}&=&<{2 \pi s \over \hbar} \int {d p dq \over (2 \pi
  \hbar)^6}
|{\cal T}^r(k,p,q)|^2 
\nonumber\\&& \times
\delta
(\varepsilon_k+\varepsilon_p-\varepsilon_{k+q}-\varepsilon_{p-q}) 
f_p(1-f_{k-q})(1-f_{p+q})>
\nonumber\\
{1\over \tilde \tau}&=&<{2 \pi s \over \hbar} \int {d p dq \over (2 \pi
  \hbar)^6}
|{\cal T}^r(k,p,q)|^2 
\nonumber\\&& \times
\delta((k-p)\cdot {q\over m}) 
f_p(1-f_p)>
\nonumber\\
\Delta_t&=&-\frac 1 \hbar <{\partial \over \partial \omega} {\rm Im}
\ln {\cal T}^r(k,p,q,\omega)|_{\omega=\varepsilon_k+\varepsilon_p}>.
\label{tau}
\ee
The local equilibrium distribution $F_0=\tilde f_0-{\tau/\tilde\tau} \tilde f_0(1-\tilde f_0)$
is specified here in such a way that equilibrium
$\tilde f_0$ is a solution of the stationary problem. Conservation laws are
enforced in the usual way as for relaxation time approximation
constructing the local equilibrium appropriately. We remark that for
Fermi ground state $\tilde f_0=(0,1)$ we have  $F_0=(0,1)$. For the
reason of simplicity we will restrict to states above Fermi level
$F_0\approx 0$. All
discussions can be performed with any given $F_0$.

One can see that the equation (\ref{2}) appears from (\ref{kine}) if 
\be
a&=&{\tau\over \tilde \tau}\qquad
\sigma=\tau \qquad
\tau_d=\Delta_t.
\label{cor}
\ee
Equations (\ref{kine}) and (\ref{tau}) are the main results of the
paper. It shows that from the nonlocal kinetic
equation for 
dense interacting Fermi systems one can derive an
evolution equation which shows the onset of chaos in that an
oscillation occurs and with increasing parameter $a$, see
(\ref{crit}), bifurcations appear and deterministic chaos sets in. 
This is due to the fact that besides the stationary solution $\tilde f_0$ of
(\ref{kine}) which corresponds to the stationary fixed point $0$ of
(\ref{2}) we have a branching fixed point which corresponds to
$1-1/a$. 
For the occupied states $F_0\approx 1$ we would get
the fixed points $(1,-1/a)$ of
(\ref{2}) which would result into a minus sign in front of the $2$ in
equation (\ref{crit}). 

\section{Model calculation}
To be specific we will now calculate the parameter (\ref{cor}) for an 
interacting Fermi system.
We use as an exploratory model the separable interaction \cite{Y59,SRS90} which is written
in terms of the difference momenta of incoming $p_1$ and outgoing particles $p_2$ 
\be
V(p_1,p_2)={(2 \pi\hbar)^3 \lambda \over m (\beta^2+p_1^2)(\beta^2+p_2^2)}
\ee
in terms of  two parameter, the coupling constant $\lambda$ and the
inverse potential range $\beta$. 
With the help of this potential Bethe-Salpeter equation or
retarded ${\cal T}$-matrix equation ${\cal T}^r=V+VG^r{\cal T}^r$
can be solved
\be
{\cal T}^r(k,p,q,\omega)&=&{(2 \pi\hbar)^3 \lambda /m \over (\beta^2+(k-p)^2/4)(\beta^2+(k-p+2
  q)^2/4)}\nonumber\\&&\times 
(1+{\pi^2 \lambda \over \beta(\sqrt{m \hbar \omega} +i\beta)^2})^{-1}.
\ee
Calculating the scattering phase shift $\cot \delta={\rm Re}{\cal
  T}^r/{\rm Im}{\cal T}^r$ 
\be
p \cot \delta={(p^2+\beta^2)^2\over 2 \pi^2 \lambda}+{p^2-\beta^2\over
  2 \beta}=-{\hbar \over a_0}+{r_0\over 2 \hbar} p^2+..
\ee
shows that the parameter $\beta$ and $\lambda$ are linked to the
scattering length $a_0$ and the potential range $r_0$
in the following way
\be
a_0&=&{2 \hbar\over \beta} (1-{\beta^3 \over \pi^2
  \lambda})^{-1}\nonumber\\
r_0&=&{3 \hbar \over \beta} (1-{4 \hbar \over 3 a_0 \beta}).
\label{ar}
\ee
Therefore we can describe with this simple rank-one  separable
potential the scattering length $a_0$ and the range $r_0$ of the
interaction. 
Alternatively to the potential range one could fit to the
bound state energy $\omega=E_B<0$ which yields 
$\lambda=\beta(\beta-m E_B)^2/\pi^2$. We neglect here medium effects on the
${\cal T}$-matrix which can be considered as well \cite{SRS90}. 
Now we can calculate the parameter (\ref{tau}) explicitly in the low temperature
limit. The angular and energy integrals can be separated in the usual
manner \cite{SMJ89,BPE91} and we obtain in lowest order temperature $T$
\be
{1\over \tau}&=&{\pi T^2 \over 3 \hbar \epsilon_f}\,  a_s^2 \,  c_1(\beta/p_f) s
\nonumber\\
 {1\over \tilde \tau}&=&{4 T \over \pi \hbar} \, a_s^2 \,
 {\beta^2\over p_f^2} \, c_2(\beta/p_f)s
\nonumber\\
\Delta_t&=&{\hbar \over 4 \epsilon_f} c_3(a_s,\beta/p_f)
\label{23}
\ee
with the Fermi energy $\epsilon_f=p_f^2/2m$ and the dimensionless
scattering length $a_s=a_0 p_f/\hbar$. 
The functions $c_1,c_2$ describing finite potential
range effects and read
\be
c_1(x)&=&\int\limits_0^1 dy {x^8 \over (x^2+1-y^2)^4}\approx1-{8\over 3
  x^2}+o(x^{-4})
\nonumber\\
c_2(x)&=&\int\limits_{-1}^1 dy {2 \sqrt{2} x^6 \over (1\!-\!y\!+\!2 x^2)^3
  \sqrt{1\!-\!y}}\approx 1\!-\!{1\over x^2}\!+\!o(x^{-4})
\nonumber\\
c_3(a,x)&=&
\frac{a\,x^4\,\left( -6 + 3\,a\,x + 2\,x^2 \right) }
  {\left( 1 \!+\! x^2 \right)
    \left( 4x^4 + ( 2 \!-\! ax)\left(2\!-\!a x\!+\!4 x^2\!-\!4 a x^3 \right) \right
    )} 
\nonumber\\&\approx&{a\over 4( 1+a^2)} +o(x^{-2})
\label{c3}
\ee
where the first two integrals are elementary functions. 
The final result for the parameter of (\ref{2}) are now represented in
terms of the microscopic scattering length, $a_s=p_f a_0/\hbar$, and
the range of the potential (\ref{ar}) as
\be
a&=&{\epsilon_f \over T} \,{12 c_2\over \pi^2 c_1} \,{\beta^2\over p_f^2}
\nonumber\\
\sigma&=&{\hbar \epsilon_f\over T^2} \, {3 \over \pi c_1 s} \, {1\over a_s^2}\nonumber\\
\tau_d&=&{\hbar \over 4 \epsilon_f} \,c_3.
\label{cond}
\ee
From this we can now give the conditions for the onset of
bifurcation and route to chaos according to (\ref{crit}). 

First it is interesting to discuss the sign of $\tau_d$ required for
(\ref{crit}). Since $c_3$ in (\ref{c3}) determine $\tau_d$  via (\ref{23}) we obtain 
\be
\begin{array}{rcrclcr}
a_0&<&0: \ \ \tau_d&\gtrless&0 \ \ {\rm for}\ p_f &\gtrless& p_{c} 
\\&&&&&&\\
0<a_0&<&{2\hbar \over \beta}: \ \ \tau_d&\gtrless&0 \ \ {\rm for}\ p_f &\lessgtr& p_{c}
\\&&&&&&\\
a_0&>&{2\hbar \over \beta}: \ \ \tau_d&>&0 \ \  && 
\end{array}
\label{cases}
\ee
with $p_{c}={\beta /
  \sqrt{3-{3 a_0 \beta \over 2 \hbar}}}$. It shows nicely the interplay 
between the range of potential and the 
density. For lower densities than the critical one, $p_f<p_c$, the delay time
follows the sign of scattering length. For higher densities the
delay time flips the sign if the scattering length is lower than $2
\hbar/\beta$.

Now
we analyze (\ref{crit}) and (\ref{om}) for the low temperature expansion compared to the
Fermi energy. 
The final resulting condition
for the occurrence of oscillations with the frequency
$\Omega$ reads
\be
&&
\begin{array}{rcrcl}
\tau_d&<&0: \ \ T&<&\epsilon_f 
{3\pi^2 \over 2 c_2  c_3 \sqrt{1-\zeta}} 
\left ( 
{\hbar \over \beta a_0}
\right )^2 
\\&&&&\\
&&\Omega&=&{\epsilon_f \over \hbar} 
\left (
{3 \pi \over 2 c_3}+{s c_1\over 18} 
\left ({p_f a_0\over \hbar} \right)^2 
\left ({T\over \epsilon_f}\right )^2
\right )
\\ &&&&\\
\tau_d&>&0: \ \ T&>&\epsilon_f {\pi^2 \over 2 c_2
  c_3 \sqrt{1-\zeta}} \left ( {\hbar \over \beta a_0}\right )^2 
\\&&&&\\
&&\Omega&=&{\epsilon_f \over \hbar} \left (
{\pi \over 2 c_3}+{s c_1\over 6} \left ({p_f a_0\over \hbar} \right
)^2 \left ({T\over \epsilon_f}\right )^2\right )
\end{array}
%\nonumber\\&&
\label{cases1}
\ee
where $\zeta={\pi^3 c_1 \over 6 s c_3 c_2^2} \left ({\hbar p_f\over a_0
  \beta^2} \right )^2\ll 1$ according to the expansion with respect to
the range of the potential.
These conditions for the temperature together with (\ref{cases}) allow
to calculate the density and temperature range for which the
oscillations appear and chaotic behavior is possible.

In the end we should discuss the validity range of the kinetic equation.
The nonlocal kinetic equation (\ref{kin}) we have started from has been derived under the assumption that the delay time is smaller than the relaxation time. This translates into $\tau_d<\sigma$ and from (\ref{cond}) we obtain 
\be
T<\epsilon_f\sqrt{12\over \pi s c_1 c_3} {1\over a_s}
\label{hcase}
\ee
which has an overlapping region with (\ref{cases1}) and is consistent with the low temperature expansion used here. To illustrate this we want to consider nuclear matter with realistic parameter. The triplet channel has a scattering length of $a_0=5.379$fm and the deuteron bound state energy is reproduced if the parameter for the separable potential is $\beta=1.4488\hbar/$fm \cite{Y59}. Then the third case of (\ref{cases}) applies. The corresponding third case of (\ref{cases1}) together with (\ref{hcase}) gives then the criterion on the temperature $0.118 \epsilon_f>T>0.0813 \hbar \epsilon_f/$fm$p_f$ which makes only sense if the density or the Fermi momentum $p_f>2.24\hbar/$fm which corresponds to about four times nuclear matter density $n_0=0.14$fm$^{-3}$. The corresponding temperatures will be around $10$MeV and the oscillation period $1/\Omega\approx 2$fm/c. These parameter are reached in experimental heavy ion reactions. Therefore we will find a parameter range where bifurcation and the route to chaos can occur in the time evolution of the distribution function itself. 

\section{Conclusion}
Summarizing, dense interacting Fermi system are described by nonlocal
kinetic equations which can be shown to be of the type of delay
differential equations. They interpolate between
finite time stepping of the logistic  map and continuous time
derivatives. The important novel observation is that the distribution
function itself shows temporal oscillations around the equilibrium value with
a basic frequency of about the Fermi energy. For
certain density and temperature conditions, (\ref{cases1}), this oscillation bifurcates
and the route to chaotic time evolution sets in. It could be
considered as a signal of
onsetting phase transition in accordance with the analogy found in 
\cite{FKR87} between mean-field phase transitions and bifurcations of
discrete maps. 

In the end it should be commented on the entropy. It is clear that from the above described possibility of oscillations in the time evolution of the 
distribution function follows that also the entropy will oscillate. This has been seen already in the collisionless but selfconsistent Vlasov equation \cite{Mr97,MP96} 
which appears if we neglect collisions in (\ref{2}) at all and calculated the quasiparticle energy on the drift side from the selfconsistent mean field. The inclusion of the collisions are expected to damp this oscillations. As we have shown above there might be a regime where the correlations are such that oscillations pertain. In such cases the entropy would oscillate. The complete 
balance equations \cite{LSM97} which follows from the nonlocal kinetic theory (\ref{2}) show that the observables contain besides the Landau one--particle parts also correlated two-particle parts. Though the explicit forms are known for the density, current, energy and stress tensor and their corresponding conservation laws are proven \cite{LSM97}, we have not been able yet to reconstruct the explicit form of the entropy beyond the single particle expression. Therefore we could not find an explicit H-theorem with two-particle correlations included. This will remain to future work.

%The range of applications are Fermi systems
%with short range correlations such that the range of potential, (\ref{ar}) is
%smaller than the inverse Fermi-momentum and the temperature fulfills
%the condition (\ref{cases1} in terms of scattering length and range of
%potential. 

\acknowledgements
The discussions with  
            Marco Ameduri, Marek Berezowski, 
Holger Kantz, Rainer Klages and Pavel 
Lipavsk\'y are gratefully acknowledged.

%\bibliography{kmsr,kmsr1,kmsr2,kmsr3,kmsr4,kmsr5,kmsr6,kmsr7,delay2,delay3,spin,refer,chaos,gdr,micha}
%\bibliographystyle{prsty}

\end{document}